\documentclass[aps,twocolumn,nofootinbib]{revtex4}
\usepackage{graphicx}
\usepackage{dcolumn}
\usepackage{amssymb}
\usepackage{bm}
\usepackage{enumerate}

\def\be{\begin{enumerate}}
\def\ee{\end{enumerate}}
\def\beq{\begin{equation}}
\def\eeq{\end{equation}}
\def\bea{\begin{eqnarray}}
\def\eea{\end{eqnarray}}

\def\g{\gamma}

\def\d{\delta}

\def\l{\lambda}
\def\o{\omega}

 \def\half{\textstyle{\frac{1}{2}}}
\def\3halfs{\textstyle{\frac{3}{2}}} \def\em{\it}

\def\nab{\nabla}

\def\ben{\begin{enumerate}}
\def\een{\end{enumerate}}
\def\bitem{\begin{itemize}}
\def\eitem{\end{itemize}}

\def\uu{\underline{u}}

\begin{document}

\title{\bf \Large
Einstein-Aether Waves}
\author{T. Jacobson}
 \altaffiliation[Permanent address: ]
{Department of Physics University of Maryland, College
Park, MD 20742-4111, USA}
\affiliation
{Insitut d'Astrophysique de Paris, 98 bis Bvd.~Arago, 75014 Paris, FRANCE}
\author{D. Mattingly}
\affiliation{Department of Physics, University of
California-Davis, Davis, CA 95616, USA}

\begin{abstract}
Local Lorentz invariance
violation can be realized by introducing extra tensor fields in
the action that couple to matter.  If the Lorentz violation
is rotationally invariant in some frame, then it is characterized
by an ``aether'', i.e. a unit timelike vector field. General
covariance requires that the aether field be dynamical. In this
paper we study the linearized theory of such an aether coupled to
gravity and find the speeds and polarizations of all the wave
modes in terms of the four constants appearing in the most general
action at second order in derivatives.  We find that in addition
to the usual two transverse traceless metric modes, there are
three coupled aether-metric modes.

\end{abstract}

\maketitle

\section{Introduction}

Recently there has been an explosion of research on the
possibility that Lorentz invariance is violated by quantum gravity
effects (see e.g.~\cite{Kostelecky:Meeting} and references therein).
In a non-gravitational setting,  it suffices to specify fixed
background fields violating Lorentz symmetry in order to formulate
the Lorentz violating (LV) matter dynamics. However, fixed
background fields break general covariance. If we are to preserve
the successes of general relativity in accounting for
gravitational phenomena, breaking general covariance is not an
option. The obvious alternative is to promote the LV background
fields to dynamical fields, governed by a generally covariant
action. Virtually any configuration of any
matter field breaks Lorentz invariance, but this differs in an
important way from what we have in mind. The LV background fields
we are contemplating are constrained either dynamically or
kinematically not to vanish, so that every relevant field
configuration violates local Lorentz symmetry everywhere,
even in the the ``vacuum".

If the Lorentz violation preserves a three-dimensional
rotation subgroup, then the
background field must be only a timelike vector, which might be
described by the gradient of a scalar, or by a vector field. In
this paper we consider just the case where the LV field is a unit
timelike vector $u^a$, which can be viewed as the minimal
structure required to determine a local preferred rest
frame.
We call this
field the ``aether", as it is ubiquitous and determines a locally
preferred frame at every point of spacetime. Kinetic terms in the
action couple the aether directly to the spacetime metric, in
addition to any couplings that might be present between the aether
and the matter fields. We refer to the system of the metric
coupled to the aether as ``Einstein-aether theory".

Here we investigate the linearized wave spectrum of
this theory, and determine the complete set of mode speeds
and polarizations for generic values of the free
parameters in the action. (Results for different
special cases were previously published in Refs.~\cite{jacobmatt,mattjacob}.)
These results 
identify the choices of constants in the 
action for which the linearized field equations are 
hyperbolic (and hence admit an initial value formulation), 
and they will be useful in extracting
the observable consequences of such an aether field.

Related work goes back at least to the 1970's, when Nordtvedt and
Will began a study of vector-tensor theories of
gravity~\cite{nordwill,willnord,hellnord,willbook}, which differed
from the present work primarily in the fact that the norm of the
vector was not constrained. Gasperini, using a tetrad formalism,
studied in a series of papers~\cite{Gasperini} an equivalent
formulation of the Einstein-aether theory studied here. Further
related work has been done by Kostelecky and Samuel~\cite{KosSam}
and Jacobson and Mattingly~\cite{jacobmatt} in the special case
where the aether dynamics is Maxwell-like.  The spherically
symmetric weak field solutions were found for the general
Einstein-aether theory by Eling and Jacobson~\cite{Eling}.
Vector-tensor theories have been studied in a cosmological context
by Clayton and Moffatt~\cite{moffatv, moffattv2} and Bassett et
al~\cite{Bassett:2000wj}. The issues of causality and shocks in
vector-tensor theories were studied by Clayton~\cite{clayton}.
Further discussion on previous work can be found
in~\cite{jacobmatt,Eling}. A proposal for Lorentz symmetry
breaking via a scalar field with unusual kinetic term that makes
the gradient tend to a timelike vector of constant norm has
recently been investigated by Arkani-Hamed et al~\cite{Ark1,Ark2}.
Most recently, the issue of Lorentz violation in a gravitational
setting has been examined in a systematic way by
Kostelecky~\cite{KosGrav}.

\section{Einstein-aether theory}
In the spirit of effective field theory, we consider a
derivative expansion of the action for the metric $g_{ab}$ and
aether $u^a$.  The most general
action that is diffeomorphism-invariant
and quadratic in derivatives is
\begin{equation} \label{eq:action}
S=\frac {1} {16 \pi G} \int \!d^4x \sqrt{-g} \Bigl(-R
+{\cal L}_u
- \lambda (u^a u_a
-1)\Bigr)
\end{equation}
where
\begin{equation}
    {\cal L}_u=-K^{ab}{}{}_{mn} \nabla_a u^m \nabla_b u^n
\end{equation}
with
\begin{equation}
    K^{ab}{}{}_{mn}=c_1 g^{ab} g_{mn} + c_2 \delta^a_m \delta^b_n
    + c_3 \delta^a_n \delta^b_m + c_4 u^a u^b g_{mn},
\end{equation}
 $R$ is the Ricci scalar, 
and $\lambda$ is a Lagrange multiplier that enforces the unit
constraint. The metric signature is $({+}{-}{-}{-})$,
units are chosen such that c=1, and other than the
signature choice we use the conventions
of \cite{wald}.

The presence of the Lagrange multiplier and $c_4$ terms
differentiate this theory from the vector-tensor theories
considered in~\cite{willbook}.
The possible term $R_{ab} u^a u^b$ is
proportional to the difference of the $c_2$ and $c_3$
terms via
integration by parts and hence has been omitted.
We have also omitted any matter
coupling since we are interested here in the dynamics of the
metric-aether sector in vacuum.

Varying
the action (\ref{eq:action}) with respect to $u^a$, $g^{ab}$,
and $\l$ yields the field equations
\begin{eqnarray}
 \nabla_a J^a{}_m - c_4 \dot{u}_a \nabla_m u^a &=&  \lambda u_m \\
  G_{ab} &=& T_{ab} \\
  g_{ab} u^a u^b &=& 1
\end{eqnarray}
where to compactify the notation we have defined
\beq J^{a}{}_{m} = K^{ab}{}_{mn} \nabla_b u^n \eeq
and
\beq \dot{u}^m = u^a\nabla_a u^m, \eeq
and the aether stress tensor is
\begin{eqnarray}
T_{ab}&=&
\nab_m(J_{(a}{}^m u_{b)} - J^m{}_{(a} u_{b)} - J_{(ab)}u^m) \nonumber\\
&&+ c_1\, \left[(\nab_m u_a)(\nab^m u_b) -(\nab_a u_m)(\nab_b u^m) 
\right]\nonumber\\
&&+ c_4\, \dot{u}_a\dot{u}_b\nonumber\\
&&+\left[u_n(\nab_m J^{mn})-c_4\dot{u}^2\right]u_a u_b \nonumber\\
&&-\frac{1}{2} g_{ab}{\cal L}_u.
\label{aetherT}
\end{eqnarray}
In the above expression the constraint has been used
to eliminate the term that arises
from varying $\sqrt{-g}$ in the constraint term in
(\ref{eq:action}),
and in the fourth line $\lambda$ has been eliminated using
the aether field equation.

\subsection{Linearized field equations}
The first step in finding the wave modes
is to linearize the field
equations about the flat
background solution with Minkowski metric
$\eta_{ab}$ and constant unit vector
$\underline{u}^a$. The fields are
expanded as
\bea
g_{ab}&=&\eta_{ab} + \gamma_{ab}\\
u^a &=& \uu^a + v^a.
\eea
The Lagrange multipler $\l$ vanishes in the
background, so we  use the same notation
for the its linearized version.
Indices are raised and lowered with $\eta_{ab}$.
We adopt Minkowski coordinates $(x^0, x^i)$
aligned with $\uu^a$, i.e. for which $\eta_{ab}={\rm diag}(1,-1,-1,-1)$
and $\underline{u}^a=(1,0,0,0)$. The letters
$i,j,k,l$ are reserved for spatial coordinate indices,
and repeated spatial indices are summed with
the Kronecker delta.

Keeping only first order terms in $v^a$ and $\gamma_{ab}$, the
field equations become
\begin{eqnarray}
  \partial_a J^{(1)a}{}_m &=&  \lambda \underline{u}_m \label{eq:linu}\\
  G^{(1)}_{ab} &=& T^{(1)}_{ab}\label{linE} \\
  v^0  +\half \gamma_{00}&=&0\label{lincon}
\end{eqnarray}
where the superscript (1) denotes the first order part of
the corresponding quantity.
The linearized Einstein tensor  is
\bea
G^{(1)}_{ab} &=& -\half\square\g_{ab}-\half\g_{,ab}\nonumber\\
&&+\g_{m(a,b)}{}^m+\half\eta_{ab}(\square\g -\g_{mn,}{}^{mn}),
\eea
where $\g=\g_m{}^m$ is the trace, while the linearized aether
stress tensor is
\bea    T^{(1)}_{ab}&=&
\partial_m[J^{(1)}_{(a}{}^m \uu_{b)}  
-J^{(1)m}{}_{(a} \underline{u}_{b)} - J^{(1)}_{(ab)}
    \uu^m]\nonumber\\
    &&+ [\uu_n(\partial_m
    J^{(1)mn}) ] \underline{u}_a \underline{u}_b
\label{T1}
\eea
If we impose the linearized aether field equation (\ref{eq:linu})
then the second and last terms of this expression for
$T^{(1)}_{ab}$ cancel, yielding
\beq
    T^{(1)}_{ab}=
-\partial_0 J^{(1)}_{(ab)} +\partial_m J^{(1)}_{(a}{}^m \uu_{b)},
\label{T1tilde}
\eeq
The linearized quantity $J^{(1)}_{ab}$ is given by
\beq
{J}^{(1)}_{ab}=
c_1\nab_a u_b + c_2 \eta_{ab}\nab_m u^m
+ c_3\nab_b u_a + c_4 \uu_a\nab_0 u_b,
\eeq
where the covariant derivatives of $u^a$ are expanded
to linear order, i.e. replaced by
\beq
(\nab_a u_b)^{(1)}=(v_b+\half\g_{0b})_{,a}+\half\g_{ab,0}-\half \g_{a0,b}.
\label{linnabu}\eeq
This completes an explicit display of the linearized field equations.

The aether perturbations are coupled to metric perturbations,
due to the presence of the background aether vector $\underline{u}^a$.
Were it not for the aether background, the linearized aether
stress tensor (\ref{T1}) would vanish, and the metric would drop
out of the aether field equation, leaving all modes uncoupled.

\subsection{Gauge choice}

Diffeomorphism invariance of the action
(\ref{eq:action}) implies that
the field equations are tensorial, hence covariant
under diffeomorphisms. The linearized equations
inherit the linearized version of this symmetry.
To find the independent physical wave modes we must
fix the corresponding gauge symmetry.

An infinitesimal diffeomorphism generated by a vector field
$\xi^a$ transforms $g_{ab}$ and  $u^a$ by
\begin{eqnarray} \label{eq:diffeo}
  \d g_{ab}&=& {\cal L}_\xi g_{ab}= \nab_a \xi_b + \nab_b \xi_a,\\
\d u^a  &=& {\cal L}_\xi u^a= \xi^m\nab_mu^a -
u^m\nab_m\xi^a.
\end{eqnarray}
In the linearized context, the vector field $\xi^a$ is itself first order in
the perturbations, hence the linearized gauge transformations
take the form
\begin{eqnarray} 
    \gamma'_{ab}& =& \gamma_{ab} + \partial_a \xi_b + \partial_b \xi_a
\label{eq:difftransformg}\\
 v'^a&=& v^a  -\partial_0 \xi^a \label{eq:difftransformv}.
\end{eqnarray}

The usual choice of gauge in vacuum GR is Lorentz gauge
$\partial^a \bar{\gamma}_{ab}=0$, where
$\bar{\gamma}_{ab}=\g_{ab}-\half\g\eta_{ab}$. This gauge is chosen
because it simplifies the Einstein tensor. The residual gauge
freedom, which exploits the field equations,  further allows one
to impose $\g_{0i}=0$ and $\g=0$. In the present case, the aether
stress tensor (\ref{T1tilde}) contains multiple terms in the
derivatives of the metric perturbation and so Lorentz gauge is not
particularly helpful. Moreover, the residual gauge freedom cannot
be used to set  $\g_{0i}$ and $\g$ to zero since these do not
satisfy the wave equation.

Instead, a convenient choice
is to directly impose the four gauge conditions\footnote{Alternatively,
instead of setting $v_{i,i}$ to zero it is equally convenient for
finding the plane wave modes to set the spatial trace $\g_{ii}$ to
zero.}
\bea \g_{0i}&=&0 \label{gaugeg}\\
 v_{i,i}&=&0. \label{gaugev}
\eea
To see that this gauge is accessible,
note that the gauge variations of
$\g_{0i}$ and $v_{i,i}$ are, according to 
(\ref{eq:difftransformg}--\ref{eq:difftransformv}),
\bea
\d \g_{0i}&=&  \xi_{i,0}+\xi_{0,i}\label{gaugevarg} \\
\d v_{i,i}&=&-\xi_{i,i0}\label{gaugevarv}
\eea
Thus to achieve the gauge (\ref{gaugeg}--\ref{gaugev})
we must choose $\xi_0$ and $\xi_i$ to satisfy
equations of the  form
\bea
\xi_{i,0}+\xi_{0,i} &=& X_i\label{Xi}\\
\xi_{i,i0}&=& Y.\label{Y}
\eea
Subtracting the second equation from the
divergence of the first gives
\beq \xi_{0,ii}=X_{i,i}-Y, \eeq
which determines $\xi_0$ up to constants of integration by solving
Poisson's equation. Then $\xi_i$ can be determined up to a
time-independent field by integrating (\ref{Xi}) with respect to
time. Having made these choices of $\xi_0$ and $\xi_i$, (\ref{Xi})
holds, and the divergence of (\ref{Xi}) implies that (\ref{Y})
holds.

In the gauge (\ref{gaugeg}--\ref{gaugev})
the tensors in the aether (\ref{eq:linu})
and spatial
metric equations (\ref{linE}) take the forms
\bea
J_{ai,}{}^a&=&c_{14}(v_{i,00}-\half\g_{00,i0})\nonumber\\
&&-c_1v_{i,kk}-\half c_{13}\g_{ik,k0}-\half c_2 \g_{kk,0i}\label{Jgauge}\\
%
%
G^{(1)}_{ij} &=& -\half\square\g_{ij}-\half\g_{,ij}-\g_{k(i,j)k}\nonumber\\
&&-\half\d_{ij}(\square\g -\g_{00,00}-\g_{kl,kl})\\
%
%
T^{(1)}_{ij} &=&-c_{13}(v_{(i,j)0}+\half\g_{ij,00})-\half c_2\d_{ij}\g_{kk,00}
\eea
where we use the notation $c_{14}:=c_1+c_4$, etc.

\begin{table*}[t]
\caption{\label{modes}Wave mode speeds and polarizations
in the gauge $\g_{0i}=v_{i,i}=0$.}
\begin{ruledtabular}
\begin{tabular}{lll}
Wave mode&squared speed $s^2\rightarrow$ small $c_i$ limit
&polarization\\
\hline transverse, traceless metric &
$1/(1-c_{13})\rightarrow 1$&$\g_{12}$, $\g_{11}=-\g_{22}$\\
transverse aether & $(c_1-\half c_1^2+\half
c_3^2)/c_{14}(1-c_{13})\rightarrow c_1/c_{14}$&
$\g_{I3}=[c_{13}/s(1-c_{13})]v_I $\\
trace &
$(c_{123}/c_{14})(2-c_{14})/
\left[2(1+c_2)^2-c_{123}(1+c_2+c_{123})\right]\rightarrow
c_{123}/c_{14}$
&$\g_{00}=-2v_0$\\
&& $\g_{11}=\g_{22}=-c_{14}v_0$\\
&&$\g_{33}=(2c_{14}/c_{123})(1+c_2)v_0$
\end{tabular}
\end{ruledtabular}
\end{table*}

\section{Wave modes}
In General Relativity there are just two modes per spatial wave vector.
Since $v^a$ has
three independent degrees of freedom, we expect that
in the Einstein-aether case there will be
five modes all together.
We now determine the wave modes in the chosen gauge.

We assume a perturbation of the form
\bea
\gamma_{ab}&=&\epsilon_{ab} e^{ik_c x^c}\label{planewaveg}\\
v^a &=& \epsilon^a e^{ik_c x^c},\label{planewavev} 
\eea 
and choose
coordinates such that the wavevector is $(k_0,0,0,k_3)$. The gauge
conditions (\ref{gaugeg}--\ref{gaugev}) then imply
\bea
\epsilon_{0i}&=&0\\
\epsilon_3&=&0.
\eea
The problem is now to find the set of
polarizations $(\epsilon_{ab},\epsilon_a)$
and corresponding wave vectors $k_a$
for which the perturbation is a solution to the
field equations 
(\ref{eq:linu}--\ref{lincon}).

The 0 component of the aether field equation
(\ref{eq:linu}) is solved by definition of $\l$,
while the constraint equation (\ref{lincon}) implies
the relation
\beq
\epsilon_0=-\half\epsilon_{00}.
\label{linconpol}
\eeq
This leaves the spatial components of
the aether equation, together with the linearized Einstein
equation. It suffices to use the spatial components of the
Einstein equation, as the other components yield redundant
information (although they do provide useful algebraic
checks).

Inserting the plane wave ansatz
(\ref{planewaveg},\ref{planewavev}) into the field equations yields
\bea \hspace{-5 mm}
[A_I]& & \quad(c_{14} s^2 - c_1)\epsilon_I-\half
c_{13}s\epsilon_{I3}=0\\{}
[A_3]&& \quad c_{14}\epsilon_{00} +
c_{123}\epsilon_{33}+c_2\epsilon_{II}=0\\{}
{}\hspace{-10 mm}[E_{II}]&&\quad \epsilon_{00}+(1+c_2)s^2\epsilon_{33}+ 
\half \nonumber\\
&&\hspace{0mm}\times\left[(1+c_2+c_{123})s^2-1\right]\epsilon_{II}=0\\{}
&\hspace{-21.5mm}[E_{11}-E_{22}]&\quad\left[(1-c_{13})s^2-1\right]
(\epsilon_{11}-\epsilon_{22})=0\\{}
[E_{12}]&&\quad \left[(1-c_{13})s^2-1\right]\epsilon_{12}=0\\{}
[E_{I3}]&&\quad c_{13}\epsilon_I + (-1+c_{13})s\epsilon_{I3}=0\\{}
[E_{33}]&& \quad(1+c_2)\epsilon_{II}+c_{123}\epsilon_{33}=0, \eea
where $[A_i]$ and $[E_{ij}]$ indicate the components
of the aether and Einstein equations.
We use the notation $s=k_0/k_3$ for the wave speed
(which will be a true ``speed" only when $s^2>0$), and
the index $I$ is dedicated to the two transverse spatial directions
$I=1,2$, so that $\epsilon_{II}=\epsilon_{11}+\epsilon_{22}$
is the trace of the transverse spatial
part of the metric polarization $\epsilon_{ab}$.

We analyze the independent mode solutions assuming generic values
of the constants $c_{1,2,3,4}$. There are a total of five
modes, two with an unexcited aether which correspond to the usual
GR modes, two ``transverse'' aether-metric modes, and a fifth
trace aether-metric mode. The two modes corresponding to the
usual gravitational waves in GR are found when all polarization
components vanish except $\epsilon_{11}$, $\epsilon_{22}$ and
$\epsilon_{12}$. To avoid over-determining the speed $s$ the trace
equation $[E_{II}]$ must be identically satisfied, hence
$\epsilon_{II}=0$. Then the $[E_{11}-E_{22}]$ and $[E_{12}]$
equations yield the speed.

The two transverse aether-metric modes have nonzero polarization
components $\epsilon_I$ and $\epsilon_{I3}$, and the $[A_I]$ and
$[E_{I3}]$ equations together yield the speed and the ratio
$\epsilon_{I3}/\epsilon_{I}$. The fifth and final mode involves
only $\epsilon_0$ and 
the diagonal polarization components $\epsilon_{aa}$ (no sum
on $a=0,1,2,3$). To avoid over-determining the speed, the
difference equation $[E_{11}-E_{22}]=0$ must be identically
satisfied, hence $\epsilon_{11}=\epsilon_{22}=\epsilon_{II}/2$.
Equations $[A_3]$ and $[E_{33}]$ and the constraint
(\ref{linconpol}) then allow all polarization
components to be expressed in terms of just one, after which
$[E_{II}]$ determines the speed. The resulting mode polarizations and
speeds are displayed in Table \ref{modes}.

In the limit $c_i\rightarrow0$ the transverse traceless modes
become the usual gravitational waves, with unit speed. Note
that these modes are entirely decoupled from the
aether perturbations even when $c_i\ne0$.

The small $c_i$ limits
of the transverse aether and trace mode speeds depend on the ratios of the
constants. If $c_{2,3,4}$ vanish both speeds approach unity, but
any other value is possible. The wave speeds and non-zero polarization
components for
the special case $c_{2,3,4}=0$ were previously reported in
\cite{mattjacob} (the speed for the trace mode
is inverted there), and the Maxwell-like case $c_{13}=c_2=c_4=0$ was
analyzed in \cite{jacobmatt} (in both cases using different
gauges). In the latter case, the
transverse waves all have unit speed, while the trace mode has
zero speed, so does not exist as a propagating wave.

A peculiar special case occurs if $c_{14}=0$, since then
the aether wave speeds are generally infinite. This happens
because no time derivatives of the aether field then arise in
the field equation $\mbox{(\ref{Jgauge})}=0$.
(The more special case $c_{14}= c_{23}=0$ was shown
by Barbero and Villase\~{n}or~\cite{Barbero} to be equivalent
to general relativity via a $u^a$ dependent field redefinition of the metric.)

When the constants $c_i$ are chosen so that $s^2$ is positive
and finite for all modes, the linearized  equations are
evidently  hyperbolic. (It is not known  whether this property
extends to the nonlinear equations.)  In these cases, since the
dispersion relation $\o=sk$ is linear, $|s|$ represents the
signal propagation speed of disturbances. It is easily checked
from Eqns. (41)-(45) that the Einstein tensor has nonvanishing
components for each of the modes. (These equations display
components of  $G_{ab}-T_{ab}$, so just those of $G_{ab}$ remain
when the $c_i$ are set to zero.) Hence the modes all have
gauge-invariant, physical significance.

If $s^2$ is negative for a mode then the corresponding
frequency is imaginary, indicating the existence of 
exponentially growing and decaying solutions. In such 
a case the theory is unstable and hence presumably unphysical.

\section{Observational applications}

An important open question
is the sign of the energy of the various wave modes. To answer
this, it is necessary to first determine the expression for energy
in the linearized Einstein-aether theory, which has not yet been
done.

To compare the wave behavior of the theory
with observations 
the wave emission
from astrophysical sources
must be determined. To begin with, the analog
of the quadrupole formula would enable the decay
of binary pulsar orbits to be computed.
Note that the presence of transverse aether and trace
modes strongly suggests that dipole and monopole 
radiation will also exist and contribute to the energy loss. 

The wave emission depends of course on how the aether field
couples to matter. A direct coupling could lead to local Lorentz
violating effects which may exist but are already quite
constrained. However, even a small coupling to the matter source
might be
large enough to produce an observable effect. Even without any
direct coupling to matter, the extra modes will still be excited
through their coupling to the time dependent metric produced by
the moving matter sources.

The results obtained here for the linearized theory
should also be useful in computing the PPN parameters. 

\section*{Acknowledgements}
This work was supported in part by the NSF under grants
PHY-9800967 and PHY-0300710 at the University of Maryland, by the
DOE under grant DE-F603-91ER40674 at UC Davis, and by the CNRS 
at the Institut d'Astophysique de Paris.

\end{document}